\documentclass[10pt]{iopart}
\usepackage{epsfig}
\usepackage{supertabular}

\begin{document}
\title {Average Path Length in Complex Networks: Patterns and Predictions}
\author {Reginald D. Smith}
\address{Bouchet-Franklin Research Institute, P.O. Box 10051
,Rochester, NY 14610}
\date {January 13, 2008}
\ead {rsmith@sloan.mit.edu}

%\keywords{complex
%networks,scale free networks,small world,geodesic,shortest path,graph invariants}

\begin {abstract}
A simple and accurate relationship is demonstrated that links the
average shortest path, nodes, and edges in a complex network. This
relationship takes advantage of the concept of link density and
shows a large improvement in fitting networks of all scales over the
typical random graph model. The relationships herein can allow
researchers to better predict the shortest path of networks of
almost any size.

\end {abstract}
\maketitle
%\paragraph
The research of complex networks has exploded over the past decade
with literally thousands of papers describing and theorizing about
such networks in all details. This explosion of research followed
the widespread availability of large network databases aided by the
advance of computer technology and widespread online applications
used by millions of users. Among the most prominent and well-known
studies have been those of the Internet \cite{Faloutsos}, metabolic
pathways\cite{metabolic}, and scientific
collaborations\cite{scicollab1, scicollab2}. Other networks have
also included sexual contacts\cite{sexual}, instant
messaging\cite{instant message}, Congressional
committees\cite{congress}, jazz musicians\cite{jazzcite},
blogs\cite{sinablog}, airports\cite{chinaair}, and rappers
\cite{rappers}.

%\paragraph
Several review articles have highlighted the main features and
characteristics of complex networks \cite{review1, review2,
review3}. One of the most studied and important features of a
complex network as been found to be the average path length (or
characteristic path), $\overline{l}$ that characterizes a network.
It describes the average number of links that form the shortest path
between any two nodes in the network. This property, more than any
other, gives rise to what is known as "small world" behavior.

\section {Brief Properties of $\overline{l}$}

In their seminal work that helped ignite research into small world
phenomena, Watts and Strogatz \cite{movie2} describe small world
networks as those which are connected, where the number of nodes is
much larger than the average degree per node, and the average path
length scales with $\log{n}$. Though random graphs can exhibit small
world behavior, most graphs in the real world are not random and are
often distinguished from random graphs by a relative high degree of
clustering among nodes as measured by the clustering coefficients.
%\paragraph
Watts and Strogatz also described an estimate from random graph
theory for the average path length of a random graph, which has
become very useful for comparison with real networks,
\begin{equation}
\overline{l} \approx \frac{ln{N}}{ln{\langle{k}\rangle}}
\end{equation}

where $N$ is the number of nodes and $k$ is the average degree per
node in the network which is $E/N$ for directed networks and $2E/N$
for undirected networks where $E$ is the number of links (edges) in
the graph. This equation gives a very good approximation for many
networks and though it is not exact, it usually gives a good rough
estimate. However, as an approximation it is usually only used to
compare the average path length of a graph using real or simulated
data and a similar random graph with the same $N$ and
$\langle{k}\rangle$.

There has been much more work done on $\overline{l}$ describing its
theoretical relationship with the small world network it
characterizes
\cite{pathlength1,pathlength2,pathlength3,pathlength4,pathlength5}.
Small-world networks have been analyzed using percolation theory and
mean field theory among others to attempt to understand the exact
nature of the transition from a "large" to a small world network.
Since $\overline{l}$ is one of the key parameters that signifies
such a change, its theoretical relationship has been investigated in
order to relate it to other properties of complex networks such as
the correlation length of the network.

\section {Link Density and Average Path Length}

Though the random graph approximation is useful, it can be asked
whether there is a better model for complex networks that can
explain known data. Complex networks, despite having similar average
path lengths or clustering coefficients can vary in other measures
such as first order degree distributions, assortative (or
disassortative) mixing, and sizes of connected components. Given the
many important topological features, much less the feedback with the
dynamics on the network that affects network evolution, it can be
questioned whether any more precise generalization is possible among
complex networks.

%\paragraph
One key concept that can link many disparate graphs, despite their
number of nodes, is the concept of network density. Network density
has been described in some papers\cite{linkdensity1,wikipedia}. The
definition used here is the ratio of the number of edges in the
network over the total possible number of edges in the complete
graph
\begin{equation}
\overline{a} = \frac{2E}{N(N-1)}
\end{equation}

The network density has a maximum of 1 in a complete graph and a
minimum of $2/(N-1) \approx 2/N$ in a simple ring topology. This
density is also identical to the value of p, the probability two
nodes will be connected by a link. However, it will be referred to
as density throughout this paper since this paper does not
concentrate on aspects of probability or percolation theory. This
density will be used for both directed and undirected networks. For
simplification, since the number of nodes in a network is usually
$N\gg1$ the link density can also be approximated as

\begin{equation}
\overline{a} = \frac{2E}{N^2}
\end{equation}

However, the link density does not directly correlate with the
average path length. Matching $\overline{a}$ against $\overline{l}$
shows very little relationship. Part of the problem is that with
increasingly larger networks, a lower link density is sufficient to
obtain a given average path length. In general, a larger network has
a much smaller $\overline{a}$ for a given $\overline{l}$ than a
smaller network with a similar $\overline{l}$.

%\paragraph

Though the relationship between the two variables is tenuous, their
product has several interesting properties

\begin{equation}
D = \overline{a}\overline{l}
\end{equation}

$D$, is equal to 1 for both complete graphs
($E=N^2/2,\overline{l}=1$) and directed ring topologies
($E=N,\overline{l}=N/2$) which as strongly connected clusters are
networks with the longest possible average path. The undirected ring
topologies have $\overline{l} = N/4$. However, outside these two
extreme cases,$D$, typically does not equal 1 but has a much lower
value, but greater than 0. The value of D varies much with
$\overline{a}$ so also has a large dependence on the size of the
network with $\overline{l}$ having a minimal impact.

A way to resolve the issue is to find a method of normalizing the
network density so it is comparable across networks of all sizes. I
define a normalized network density, $\overline{a}_{s}$, by taking
the logarithm of the network density with a base of $N/2$ and adding
1 which is equivalent to
\begin{equation}
\overline{a}_{s} =1 + \frac{\log{\frac{2E}{N^2}}}{\log{N/2}}
\end{equation}
where $\log$ here designates a natural logarithm. When the network
is a complete network the normalized network density is 1, while for
ring topology, it equals 0. Therefore, the size of the network will
not affect the minimum network density.

Not only does this normalized network density allow you to compare
network densities over networks of various sizes, it actually
demonstrates a correlation with the path length of the network, in
particular the inverse of $\log{\overline{l}}$.
%\paragraph

\begin{figure}

    \centering
    \includegraphics[height=2in, width=2in]{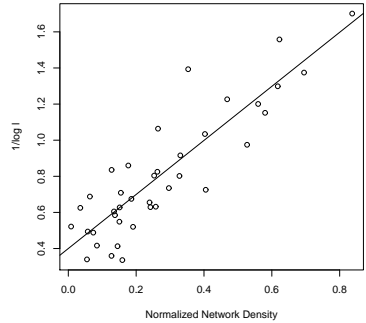}
        \caption{Plots of $\frac{1}{\log{\overline{l}}}$ vs. normalized network density. The slope
        of the fit is 1.5 respectively with an intercept of 0.4. $R^2$ of 0.78}
    \label{mainplots}
\end{figure}

The graph in Figure \ref{mainplots} was developed using data from 39
different networks described in various papers. The values of these
networks are shown in Appendix I. These networks are of many
different types and have been given broad categorizations following
those used by Newman\cite{review1}.

\begin{figure}

    \centering
    \includegraphics[height=2in, width=2in]{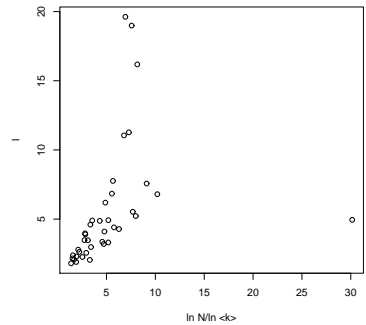}

        \caption{Plots of the same network data with $\overline{l}$ vs.
        $\frac{\ln{N}}{\langle{k}\rangle}$. The data fits relatively well for
        graphs with small average path lengths or small N but
        shows greater disparities when these conditions are not
        met.}
    \label{randomgraph1}
\end{figure}

This relationship was experimentally discovered and not quite
expected. In fact, one of its more interesting properties is how it
fits disparate networks, of all scales and average path lengths,
accomodating the data better than the random graph estimation in
Figure \ref{randomgraph1}
%\paragraph

In fact in Figure \ref{mainplots} the main points that do not fit
well to the least squares line are those from biological networks,
including the food webs (freshwater and marine) and metabolic
networks. This may indicate either the data on these networks or
incomplete or the underlying organizational property driving this
relationship is less active in biological networks. The relation
derived from the regressions implies

\begin{equation}
\label{fitequation}
 m\overline{a}_{s}+C = \frac{1}{\log{\overline{l}}}
\end{equation}

this allows us to relate $\overline{l}$ to the normalized density
with the equation

\begin{equation}
\label{lengthequation} \overline{l} = e^{\frac{1}{m\overline{a}_s
+C}}
\end{equation}

A quick but interesting example can be made using equation
\ref{lengthequation}. Assuming the US population is 300M and
accepting Milgram's six degrees of separation ($\overline{l} = 6$)
we can estimate the average $\langle{k}\rangle$ for the US
population at 14.6. This is much less than the 25.9 estimated from
random graph theory (and assumes us to be substantially less
gregarious).

%\paragraph

A key question about the relationship is Figure \ref{mainplots} is
how widely it applies to all types of networks. All of the networks
sampled are described by authors as having "scale-free" or
"long-tailed" characteristics. Obviously, graph theory does not
constrain a network from being of this type so by looking at the
relationship using data from an artificial random graph we can begin
to push the boundaries of its applicability.

%\paragraph
\begin{figure}

    \centering
    \includegraphics[height=2in, width=2in]{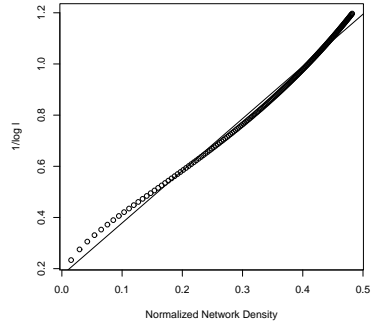}

        \caption{Plots of a graph of 1,000 nodes from 1,100 to 20,000 nodes where $\overline{l}$
        = $\frac{\ln{N}}{\ln{\langle{k}\rangle}}$}
    \label{randomgraph2}
\end{figure}
In Figure \ref{randomgraph2} it is clearly visible that the linear
relation among real networks also holds for random graph data.

The slope of the plot from a random graph approximation is 2.04
which is slightly steeper than the slope of data from real networks.
Therefore, this relationship is likely widely held among many
small-world networks with a variety of topologies though there are
likely exceptions.

\section{Discussion}

First, it should be acknowledged that though this relationship seems
to fit a wider variety of real networks than random graph theory, it
is not perfect. From the standards of theoretical prediction, the
statistical fit still allows much leeway for the relationship
between the quantities plotted against each other. However, this
relationship does fit data more consistently over all size scales in
real networks than the usual random graph theory treatment.

Despite the interesting relationship this data reveals, it also
raises the question of what the parameters of the linear plot
actually mean. One clue can be gleaned from looking at the rate of
change of the average shortest path vs. the normalized link density
$\frac{\partial\overline{l}}{\partial\overline{a}_{s}}$.

Given equation \ref{fitequation} you can easily deduce from the fact
that
\begin{equation}
\label{rate} \frac{\partial\overline{l}}{\partial\overline{a}_{s}} =
-m\overline{l}[\log{\overline{l}}]^2
\end{equation}

So the slope, $m$, can be seen as the constant of proportionality
between the rate of change with increasing network density in the
average shortest path and the average shortest path. In fact,
Equation \ref{rate} gives quite intuitive solutions since $m>0$. The
larger $\overline{l}$ is, the more rapidly you can reduce the
average shortest path of the network by increasing the network
density. This intuitively fits with the observation by Watts and
Strogatz \cite{movie2} that in relatively sparse topologies,
shortcuts can drastically reduce the average path length of the
network leading to small world behavior. As the network becomes more
dense, such short cuts give incrementally smaller reduction in the
network average path. When you reach a complete network at
$\overline{l} = 1$ there is a fixed point given you have maximum
density and can no longer reduce the diameter of the network.

%\paragraph

Additionally, $m$ could be some measure of a quantity such as the
"mass" of the network. If $\log\overline{l}$ can somehow be seen as
a length then this gives additional meaning to the normalized link
density. Here $m$ would be a characteristic mass in all networks
that is distributed over a one-dimensional interval determined by
$\log\overline{l}$ and $\overline{a}_{s}$ measures the resultant
length density. However, the question of what $m$ really is as far
as a value is still unanswered. The fact that it seems consistent
across such a wide variety of networks suggests it is some constant,
perhaps of a transcendental number or ratio. This is all speculative
though. Until a firm theoretical underpinning for the above results
is made, the exact value of $m$ is still subject to speculation.

%\paragraph
In addition, although this relation seems to hold across a wide
variety of networks, there are obviously situations where equations
such as equation \ref{lengthequation} break down. For example, when
$\overline{a}_{s}$ is 0, a complete graph, $\overline{l}$ should
have a value of 1, however, this does not necessary flow from the
relations shown here. The only exception is equation \ref{rate} that
shows a fixed point at $\overline{l}=0$ as is expected. Therefore,
in the regions of nearly complete graphs or sparse graphs, possibly
where $p<p_{c}$ where $p_{c}$ is the critical probability from the
percolation theory, this relationship does not reliably apply.
However, these regions are not the domain of almost all real
networks.

\section{Conclusion}
This paper has shown that there is an intrinsic relationship between
the average path length $\overline{l}$ and the normalized link
density, related to the number of nodes and edges, that is present
in all networks. This relationship fits well in both real networks
which often have a scale-free or non-random character, but can also
describe random networks as well and likely most small world
networks. Given the breakdown of the theory near the complete and
ring topologies it may be surmised this only applies to graphs with
small-world character that have a link probability $p_{c} < p < 1$.
Much more research is needed, however, to determine the exact reason
for this relationship and in particular, the meaning of the
parameter m.

\section{Appendix I - Measures of Real Networks}

\begin{table}[!p]
\begin{center}

 \caption{Real Network Data Used in Paper}
\begin{tabular}{|p{100pt}|c|c|c|c|c|c|c|}
\hline
Network&$n$&$M$&$z$&$\overline{\ell}$&$Category$&$Type$&$Source$\\
\hline
Protein Interaction&2115&2240&2.1&6.8&biological&Undirected&\cite{protein}\\
Metabolic Network&765&3686&9.6&2.56&biological&Undirected&\cite{metabolic}\\
Neural Network&307&2359&7.7&3.97&biological&Directed&\cite{neuralnet}\\
Marine Food Web&135&598&4.4&2.05&biological&Directed&\cite{marinefood}\\
Freshwater Food Web&92&997&10.8&1.9&biological&Directed&\cite{freshfood}\\
World Trade&179&7697&86.0&1.8&economic&Undirected&\cite{worldtrade}\\
WWW Altavista&203549046&2130000000&10.5&16.18&information&Directed&\cite{wwwaltavista}\\
Wikipedia&434000&8500000&19.6&4.9&information&Directed&\cite{wikipedia}\\
WWW nd.edu&269504&1497135&5.6&11.27&information&Directed&\cite{wwwnd}\\
Sina.com Blogosphere&200339&1803051&9.0&6.84&information&Directed&\cite{sinablog}\\
Thesaurus&1022&5103&5.0&4.87&information&Directed&\cite{Thesaurus}\\
Cyworld (Korean social networking site)&12048146&190589667&31.6&3.2&social&Undirected&\cite{cyworld}\\
Biology Coauthors&1520521&11803064&15.5&4.92&social&Undirected&\cite{scicollab1,scicollab2}\\
Film Actors&449913&25516482&113.4&3.48&social&Undirected&\cite{movie1,movie2}\\
Mixi (Japan social networking site)&360802&1904641&10.6&5.53&social&Undirected&\cite{mixi}\\
Math Coauthors&253339&496489&3.9&7.57&social&Undirected&\cite{mathnet1,mathnet2}\\
Email Messages&59912&86300&1.4&4.95&social&Directed&\cite{emailmessage}\\
Physics Coauthors&52909&245300&9.3&6.19&social&Undirected&\cite{scicollab1,scicollab2}\\
Nioki.com (Instant Messaging)&50158&240758&9.6&4.1&social&Undirected&\cite{instant message}\\
Pussokram (online dating community)&29341&174662&6.0&4.4&social&Directed&\cite{pussokram}\\
Email Address Books&16881&57029&3.4&5.22&social&Directed&\cite{email}\\
Company Directors&7673&55392&14.4&4.6&social&Undirected&\cite{director1,director2}\\
Marvel Comic Characters&6486&168267&51.9&2.63&social&Undirected&\cite{comics}\\
Brazil Pop&5834&507005&173.8&2.3&social&Undirected&\cite{brazilpop}\\
Rapper Collaboration&5533&57972&21.0&3.9&social&Undirected&\cite{rappers}\\
Roman/Greek Myth Characters&1637&8938&5.5&3.47&social&Directed&\cite{myths}\\
Jazz Collaboration&1275&38326&60.1&2.79&social&Undirected&\cite{jazzcite}\\
News Article Topics&459&2763&12.0&2.98&social&Undirected&\cite{newsarticles}\\
Dolphin Network&64&159&5.0&3.36&social&Undirected&\cite{dolphin}\\
Electronic Circuits&24097&53248&4.4&11.05&technological&Undirected&\cite{electric}\\
Internet AS&10697&31992&6.0&3.31&technological&Undirected&\cite{internet}\\
Power Grid&4941&7594&3.1&18.99&technological&Undirected&\cite{movie2}\\
Warsaw Public Transport&1530&4406&5.8&19.62&technological&Undirected&\cite{polishSubway}\\
Peer to Peer Networks (Gnutella)&880&1296&2.9&4.28&technological&Undirected&\cite{Gnutella1,Gnutella2}\\
Indian Railways&587&19603&66.8&2.16&technological&Undirected&\cite{indiarail}\\
Ammonia Reaction Process&505&759&3.0&7.76&technological&Undirected&\cite{ammonia}\\
Austria Airports&133&1518&11.4&2.383&technological&Directed&\cite{austriaair}\\
China Airports&128&2304&36.0&2.07&technological&Undirected&\cite{chinaair}\\
India Airports&79&442&5.6&2.26&technological&Directed&\cite{indiaair}\\

\hline
\end{tabular}
\end{center}
\label{networks}
\end{table}


\begin{thebibliography} {1}
\bibitem{Faloutsos} Faloutsos, M., Faloutsos, P., and Faloutsos, C., Computer
Communications Review \textbf{29}, 251–262 (1999)
\bibitem{metabolic} Jeong, H., Tombor, B., Albert, R., Oltvai, Z. N.,
and Barab\'{a}si, A.-L., Nature \textbf{407}, 651–654 (2000)
\bibitem{scicollab1} Newman, MEJ., Phys. Rev. E \textbf{64},016131 (2001)
\bibitem{scicollab2} Newman, MEJ., Phys. Rev. E \textbf{64},016132 (2001)
\bibitem{sexual} Liljeros, F., Edling, C. R., Amaral, L. A. N., Stanley,
H. E., and Aberg, Y., Nature \textbf{411}, 907-908 (2001)
\bibitem{instant message} Smith, RD cond-mat/0206378 (2002)
\bibitem{congress} Porter MA, Mucha PJ, Newman MEJ, Warmbrand CM Proc. Natl. Acad. Sci. USA \textbf{102}, 7057 (2005)
\bibitem{jazzcite} P. Gleiser and L. Danon, Adv. Complex Syst. \textbf{6}, 565 (2003)
\bibitem{sinablog} Fu, F., Liu, L., Yang, K., and Wang, L., preprint math/0607361 (2006)
\bibitem{chinaair} Li, W. \& Cai, X., Phys Rev E., \textbf{69}, 046106 (2004)
\bibitem{rappers} Smith, R. J. Stat. Mech. P02006 (2006)
\bibitem{review1} Newman, MEJ  Siam Rev. \textbf{45}: 167-256
(2003)
\bibitem{review2}  Dorogovtsev SN, Mendes JFF Adv. Phys. \textbf{51}:
1079-1187 (2002)
\bibitem{review3}  Barab\'{a}si, A.L. and Albert, R, Rev. Mod. Phys.
\textbf{74}, 47, (2002)
\bibitem{movie2}Watts, D. J. and Strogatz, S. H., Nature \textbf{393}, 440–442 (1998)
\bibitem{pathlength1} Dorogovtsev, S. N. and Mendes, J. F. F., Europhys. Lett. \textbf{50}, 1–7 (2000).
\bibitem{pathlength2}Newman, M. E. J., Moore, C., and Watts, D. J., Phys. Rev. Lett.
\textbf{84}, 3201–3204 (2000)
\bibitem{pathlength3} Lochmann, A., Requardt, M. Journ. Stat. Phys. \textbf{122}, 255 (2006)
\bibitem{pathlength4} Barth´el´emy, M. and Amaral, L. A. N., Phys. Rev. Lett. \textbf{82},
3180–3183 (1999)
\bibitem{pathlength5} Almaas, E., Kulkarni, R. V., and Stroud, D., Phys. Rev. Lett.
\textbf{88}, 098101 (2002).
\bibitem{linkdensity1} Garlaschelli, D. \&  Loffredo, M. Phys. Rev. Lett. \textbf{93}, 268701 (2004)
\bibitem{wikipedia} Zlatic, V., Bozicevic, M., Stefancic, H., \& Domazet, M. Phys. Rev. E \textbf{74}, 016115 (2006)
\bibitem{protein}Jeong, H., Mason, S., Barab´asi, A.-L., and Oltvai, Z. N., Nature \textbf{411}, 41–42 (2001)
\bibitem{neuralnet}White, J. G., Southgate, E., Thompson, J. N., and Brenner, S., Phil. Trans. R. Soc. London \textbf{314},1–340 (1986)
\bibitem{marinefood}Huxham, M., Beaney, S., and Raffaelli, D., Oikos \textbf{76}, 284–300 (1996)
\bibitem{freshfood}Martinez, N. D.,  Ecological Monographs \textbf{61}, 367–392 (1991)
\bibitem{worldtrade}Serrano, M. \& Boguna, M. Phys Rev E \textbf{68}, 015101 (2003)
\bibitem{wwwaltavista}Broder, A., Kumar, R., Maghoul, F., Raghavan, P., Rajagopalan, S., Stata, R., Tomkins, A., and Wiener, J., Computer Networks \textbf{33}, 309 (2000)
\bibitem{wwwnd} Albert, R., Jeong, H., and Barab´asi, A.-L.,  Nature \textbf{401}, 130–131 (1999)
\bibitem{Thesaurus}Knuth, D. E., \textit{The Stanford GraphBase: A Platform for Combinatorial
Computing, Addison-Wesley, Reading, MA}(1993)
\bibitem{cyworld} Ahn, Y., Han, S., Kwak, H.,Moon, S., \& Jeong, H. "Analysis of Topological Characteristics
of Huge Online Social Networking Services", \textit{Proceedings of
the 16th international conference on World Wide Web}, 835, (2007)
\bibitem{movie1} Amaral, L.A.N., Scala, A., Barth\'{e}l\'{e}my, M., and
Stanley, H.E., Proc. Nat. Acad. Sci. USA \textbf{97}, 11149, (2000)
\bibitem{mixi} Yuta, K., Ono, N., \& Fujiwara, Y. preprint physics/0701168 (2007)
\bibitem{mathnet1}de Castro, R. and Grossman, J. W., Mathematical Intelligencer \textbf{21}, 51–63
\bibitem{mathnet2} Grossman, J. W. and Ion, P. D. F., Congressus Numerantium \textbf{108}, 129 (1995)
\bibitem{emailmessage}Ebel, H., Mielsch, L.-I., and Bornholdt, S., Phys. Rev. E \textbf{66}, 035103 (2002)
\bibitem{pussokram} Holme, P, Edling, C. \&, Liljeros, F. Social Networks, \textbf{26} (2), 155 (2004)
\bibitem{email} Newman, M. E. J., Forrest, S., and Balthrop, J., Phys. Rev. E \textbf{66}, 035101 (2002)
\bibitem{director1}Gerald F. Davis, Mina Yoo, and Wayne E. Baker,  Strategic
Organization \textbf{1}, 301 (2003)
\bibitem{director2}Newman, M. E. J., Strogatz, S. H., and Watts, D. J.,
Phys. Rev. E \textbf{64}, 026118 (2001)
\bibitem{comics} R. Alberich, J. Miro-Julia, F. Rossello,cond-mat/0202174
(2002)
\bibitem{brazilpop}  Silva DDE, Soares MM, Henriques MVC, et al, Physica A \textbf{332}, 559 (2004)
\bibitem{myths} Choi, Y. and Kim, H. preprint physics/0506142 (2005)
\bibitem{newsarticles} Ozgur, A. and Bingol, H. "Social Network of
Co-Occurence in News Articles", Lecture Notes in Computer Science,
\textbf{3280} (Proceedings of the 19th Annual International
Symposium on Computer and Information Sceince) 688, 2004
\bibitem{dolphin} Lusseau, D., Evolutionary Ecology, \textbf{21} (3), 357 (2007)
\bibitem{electric} Ferrer i Cancho, R., Janssen, C., and Sol´e, R. V., Phys. Rev. E \textbf{64}, 046119 (2001)
\bibitem{internet}Chen, Q., Chang, H., Govindan, R., Jamin, S., Shenker, S. J., and Willinger, W., "The origin of power laws in Internet topologies revisited", in Proceedings of the 21st Annual Joint Conference of the IEEE Computer and Communications Societies, IEEE Computer Society (2002)
\bibitem{polishSubway} Sienkiewicz, J. \& Holyst, J. Phys. Rev. E, \textbf{72}, 046127 (2005)
\bibitem{Gnutella1} Ripeanu, M., Foster, I., and Iamnitchi, A., IEEE Internet Computing \textbf{6}, 50–57 (2002)
\bibitem{Gnutella2} Adamic, L. A., Lukose, R. M., Puniyani, A. R., and Huberman, B. A.,  Phys. Rev. E \textbf{64}, 046135 (2001)
\bibitem{indiarail}Sen, P., Dasgupta, S., Chatterjee, A., Sreeram, P. A., Mukherjee, G., \& Manna, S. S., Phys Rev E \textbf{67},
036106 (2003)
\bibitem{ammonia}  Jiang, Z.,  Zhou, W.,  Xu, B., \& Yuan, W., AIChE Journal \textbf{53}, 423-428 (2007)
\bibitem{austriaair} Han, D., Qian, J., \& Liu, J. preprint physics/0703193 (2007)
\bibitem{indiaair} Bagler, G. preprint cond-mat/0409773 (2004)

\end{thebibliography}
\end{document}